\documentclass[epj]{svjour}
\usepackage{epsfig}
\begin{document}
%
%\title{%\hfill{\tiny FZJ-IKP-TH-2008-06}\\
\title{Charmed meson rescattering in the reaction {\boldmath$\bar{p}d \to \bar DD N$}}
\author{J. Haidenbauer\inst{1}, G. Krein\inst{2}, Ulf-G. Mei{\ss}ner\inst{1,3},
and A. Sibirtsev\inst{1,3}}

\institute{
Forschungszentrum J\"ulich, Institut f\"ur Kernphysik,
D-52425 J\"ulich, Germany \and
Instituto de F\'{\i}sica Te\'{o}rica, Universidade Estadual
Paulista,
Rua Pamplona, 145 - 01405-900 S\~{a}o Paulo, SP, Brazil
\and
Helmholtz-Institut f\"ur Strahlen- und Kernphysik (Theorie),
Universit\"at Bonn, Nu\ss allee 14-16, D-53115 Bonn, Germany
}
\date{Received: date / Revised version: date}

\abstract{We examine the possibility to extract information about the
${D}N$ and ${\bar D}N$ interactions from the ${\bar p}d{\to}D^0 {D^-}p$ reaction. 
We utilize the notion that the open-charm mesons are first produced in 
the annihilation of the antiproton on one nucleon in the deuteron and 
subsequently rescatter on the other (the spectator) nucleon. The latter process is 
then exploited for investigating the $DN$ and ${\bar D}N$ interactions.
We study different methods for isolating the contributions from the 
$D^0p$ and ${D^-}p$ rescattering terms.
}


\authorrunning{J. Haidenbauer et al.}
\titlerunning{Charmed meson rescattering in the reaction {\boldmath$\bar{p}d \to \bar DD N$}}

\maketitle

\section{Introduction}
\label{sec:intro}

The distortion of charm in nuclear matter remains an heavily discussed issue
since the first proposals~\cite{Matsui:1986dk} to use charmonia and open 
charm as a probe of the early stage of heavy-ion collisions, 
for a review see \cite{Satz:2005hx}. It was
expected~\cite{Batsouli} that charmed final-state interactions (FSI) either at the 
partonic or the hadronic rescattering level would not distort the spectra initially 
produced in heavy-ion collisions, because the cross sections for any such
(elastic and inelastic) scattering processes are sufficiently small. 
Furthermore, \, gluon radiation \, or
bremsstrahlung~\cite{Djordjevic1,Armesto,Djordjevic2}, which distorts 
the original charm spectrum as well, becomes the dominant energy loss mechanism
only if the heavy charmed quarks are ultra-relativistic. That is similar to the
brems\-strahlung losses of electrons passing through a hydrogen target~\cite{PDG,Moore}.
However, in the present experiments a large fraction of the heavy quarks are 
produced with momenta less than their mass and therefore the radiation losses 
might be negligible. In that case the heavy charmed quarks and antiquarks and, 
specifically, the finally detected $D$ and ${\bar D}$ mesons are presumably not
distorted in the nuclear environment and thus can probe the initial stage of the
interaction, possibly, the Quark Gluon Plasma.

On the other hand it was argued~\cite{Moore,Hees1,Hees2,Wicks} that the two basic
processes involved in the energy-loss mechanism of charmed particles  
moving in nuclear 
matter, namely  gluon radiation and elastic scattering, might be equally
important and non-negligible. Only recently the situation chan{\-}ged due to the
PHENIX and STAR experiments~\cite{Adler,Bielcik,Zhang,Zhong} at the 
Relativistic Heavy Ion Collider (RHIC). These new measurements indicate a
substantial suppression of the
production of open-charm mesons with transverse momenta above $\simeq$1~GeV/c
from central $Au{+}Au$ collisions, as compared to that from $d{+}Au$
collisions. This observation could not be assigned to  gluon radiation of
charm and thus indirectly points to an importance of distortions due to 
the FSI.

Apparently, the interactions of charmed quarks or antiquarks in nuclear matter are not
the same as the interactions of open-charm mesons. However, it is clear that a
reasonable understanding of elastic scattering involving particles with charm
on a hadronic level is highly important. While one could not measure directly 
the interaction of the charmed and
light quarks and antiquarks, the $DN$ and ${\bar D}N$ interactions can be
studied experimentally. These could serve as a basis to construct 
phenomenologically the charmed FSI at the partonic rescattering level.

The basic problem is the complete lack of relevant experimental data. This
situation is expected to change with the operation of the Facility for Antiproton 
and Ion Research (FAIR) at Darmstadt (Germany). The Proton ANtiproton at
DArmstadt (PANDA) Collaboration~\cite{Panda} intends to investigate the
distortion of open-charm mesons \cite{Hartmann,Brinkmann,Kuhn} in {\it matter}
and in the {\it vacuum}. The {\it matter} measurements are based on $D$ and $\bar
D$ meson production in antiproton annihilation on different nuclei in order to
study the $A$-dependence. The {\it vacuum} measurements explore the production of 
open-charm mesons by annihilating antiprotons on the deuteron and, through the
rescattering of the produced $D$ and $\bar D$ mesons on the spectator nucleon,
the interaction in the $DN$ and ${\bar D}N$ systems.

In the present paper we 
examine the possibility to extract information about the $DN$ and ${\bar D}N$
interactions from the ${\bar p}d{\to}D^0 {D^-}p$ reaction. The study is
based on the notion that those open-charm mesons are first produced by 
annihilating the antiproton on one of the nucleons in the deuteron and 
subsequently rescatter on the other (the spectator) nucleon. 
The latter process is then exploited for investigating the $DN$ and ${\bar D}N$ 
interactions.

To explore the potential of pertinent experiments we perform concrete 
calculations taking into account the nucleon exchange Born diagram, 
corresponding to the elementary $\bar NN \to \bar DD$ annihilation process, 
cf. Fig.~\ref{diag}a),
as well as rescattering diagrams involving the $DN$ and the ${\bar D}N$ 
interactions, see Fig.~\ref{diag}b). For those interactions we employ realistic
${\bar D}N{\to}{\bar D}N$~\cite{Haidenbauer} and $DN{\to}DN$~\cite{DN} 
scattering amplitudes. 
This is a substantial improvement as compared to previous
studies~\cite{Sibirtsev:1999js,Cassing:1999wp,Sib01} which relied on
rather simple assumptions as far as the ${D}N$ and $\bar D N$ 
scattering amplitudes were concerned. 

The paper is structured in the following way: 
In the subsequent section we introduce briefly the formalism used 
for calculating the reaction ${\bar p}d{\to}D^0 {D^-}p$. 
The utilized interaction models in the ${\bar D}N$ and $DN$ channels are
introduced and discussed in Sect.~\ref{sec:interaction}. In particular,
we present results for total and differential cross sections for
the various charge channels. 
In Sect.~\ref{sec:annihilation} a short overview on our present knowledge
on the elementary $\bar NN \to \bar DD$ reaction is given.
Our results for the reaction ${\bar p}d{\to}D^0 {D^-}p$ are shown 
in Sect.~\ref{sec:pd}. Here the emphasis is put on the exploration of
different methods for detecting and isolating the contributions from the 
$D^0p$ and ${D^-}p$ rescattering terms. For that purpose
we consider the spectator momentum distribution, Dalitz plots, the
missing mass of the exchanged meson and correlations between properly
defined scattering planes. The paper ends with a brief summary. 
As a test of our approach we also apply it to multipion production 
in ${\bar p}d$ annihilation and we compare our results with data available 
for the ${\bar p}d{\to}\pi^+2\pi^-{p}$ and ${\bar p}d{\to}2\pi^+3\pi^-{p}$ 
reactions. Those results are included in an appendix. 
 
\section{Formalism}
\label{sec:formal}

In this section we introduce the formalism we use to investigate the effects of
the $DN$ and $\bar D N$ interactions in antiproton annihilation on the deuteron. 

\begin{figure}[t]
\vspace*{-6mm}
\centerline{\hspace*{-4mm}\psfig{file=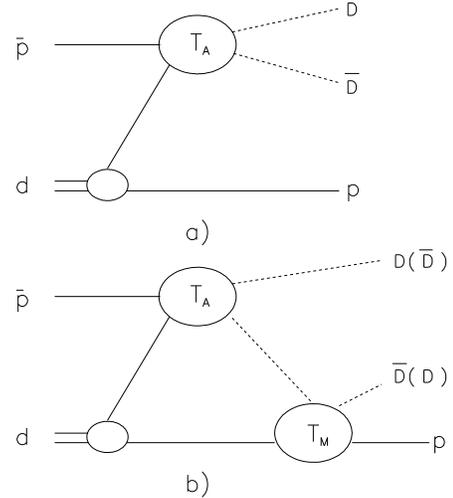,width=7cm}}
\vspace*{-7.5mm}
\caption{Contributions to the reaction ${\bar p}d{\to}D{\bar D}{N}$: 
a) The Born (nucelon exchange) diagram. ${T}_A$ denotes the annihilation amplitude. 
b) Meson rescattering diagram.
${T}_M$ denotes the meson-nucleon scattering amplitude. Note that 
both $DN$ and ${\bar D}N$ scatterings contribute to the reaction 
amplitude.
}
\label{diag}
\end{figure}

Fig.~\ref{diag} illustrates processes contributing to the reaction
${\bar p}d{\to}D{\bar D}{N}$. The diagrams of 
interest are the nucleon exchange Born diagram, Fig.~\ref{diag}a), and the
meson rescattering diagram, Fig.~\ref{diag}b). The corresponding amplitudes
will be denoted below by $T_a$ and $T_b$, respectively. 

In what follows we assume for convenience that the spectator nucleon -- the
nucleon that does not enter the annihilation vertices -- is the proton. The case
of a neutron as spectator can be treated in a similar manner. The nucleon
exchange Born diagram, Fig.\ref{diag}a), leads to the well-known reaction
amplitude~\cite{Locher0,Kolybasov1,Laget1}
\begin{eqnarray}
{  T}_a = \psi_d(p_s) \, {  T}_A,
\end{eqnarray}
where $\psi_d(p_s)$ is the momentum-space deuteron wave function, with ${\vec p_s}$
being the proton spectator momentum, and $T_A$ is the amplitude for the 
${\bar p}n \to D^-D^0$ annihilation process. After summation over spin
states the squared amplitude is given as
\begin{eqnarray}
|{  T}_a|^2 =[u(p_s)^2 + w(p_s)^2] |{  T}_A|^2,
\label{tree}
\end{eqnarray}
where $u$ and $w$ stand for the $s$- and $d$-wave components of the deuteron
wave function. The overall size of the nucleon exchange contribution
in Fig.~\ref{diag}a) is determined by the annihilation amplitude $T_A$, which
depends on the meson channels produced in the ${\bar p}n$ annihilation. 
On the other hand, the spectator momentum distribution is 
governed predominantly by the deuteron wave function, as
indicated in Eq.~(\ref{tree}), provided that 
$T_A$ is a slowly varying function of the energy. 

Next we consider the rescattering diagram of Fig.~\ref{diag}b) where one of the
mesons produced at the annihilation vertex is scattered off the spectator nucleon. 
In general, the intermediate meson in this diagram is not necessarily the same 
as the final rescattered meson. It could be an intermediate $D^*$ vector-meson,
for example. Thus, in principle a sum over all possible intermediate states is 
required. But in the following let us regard explicitly the features of the 
rescattering mechanism involving
elastic $Mp{\to}Mp$ scattering only. Note that the formalism can be easily extended 
to other, non-diagonal transitions. We average over the spins in the annihilation and
scattering vertices and take into account only the $s$-wave component $u(p_s)$
of the deuteron wave function. The $d$-wave component is expected to play a 
much less important role for the rescattering contribution \cite{Laget1,Fasano2} 
and, therefore, we ignore it here in this exploratory study. 
The integration for the rescattering diagram
runs over the three momentum of the spectator proton in the loop and can be
split into on-shell and off-shell parts, which we denote by $T^{on}_b$ and
$T^{off}_b$. The on-shell part is defined by
taking the intermediate proton to be on-shell and is given
as~\cite{Laget1,Laget2,Locher2,Fasano1,Voronov}
\begin{eqnarray}
T^{on}_b = -\frac{i}{32\pi |{\vec p}|} \int\limits_{q_{-}}^{q_{+}}\!\!
dq\, u(q) \frac{q}{E_q}\,  {  T}_M \, \, {  T}_A,
\label{rescat}
\end{eqnarray}
where $T_M$ is the meson-proton scattering amplitude, $q$ is the
internal proton loop momentum, $E_q = (q^2+m_p^2)^{1/2}$ with $m_p$ the
proton mass, and ${\vec p}$ is the sum of the momenta of the final proton and
the rescattered meson. 
The limits of the integral are given by~\cite{Byckling}
\begin{eqnarray}
q_{\pm} = \frac{|\vec p|}{s^{1/2}_{Mp}} \, E^\ast  \pm \frac{E}{s^{1/2}_{Mp}} \,
p^\ast,
\end{eqnarray}
where $E{=}E_M{+}E_p$ is the sum of the energies of the final proton and 
the rescattered meson, while $s^{1/2}_{Mp} = (E^2{-}p^2)^{1/2}$ is their 
invariant energy and
\begin{eqnarray}
{p^\ast}^2 &=& \frac{(s_{Mp}-m_p^2-m_M^2)^2-4m_p^2m_M^2}{4s_{Mp}} ,
\nonumber \\
E^\ast &=& \sqrt{{p^\ast}^2 +m_p^2},
\end{eqnarray}
with $m_M$ the meson mass. The evaluation of ${ T}_b$ requires the 
knowledge
of both amplitudes ${  T}_M$ and ${  T}_A$ within the range of $q_-$ to
$q_+$ allowed for the considered reaction.

The effect of 
the off-shell part of the rescattering integral was investigated in detail in
Refs.~\cite{Laget1,Laget2,Locher2,Fasano1,Laget3,Locher3}. In those studies 
it was found that the shape of the spectator momentum distribution,
as given by the on-shell part, remains essentially unchanged when the 
off-shell contribution is added.  At the same time, the magnitude of the 
off-shell contribution is significant and can lead to modifications of the 
on-shell results in the order of 30\% or more, 
but depends strongly on the specific off-shell behaviour of the 
annihilation and scattering amplitudes. Explicitly, the off-shell part of 
the amplitude can be written as~\cite{Laget1,Fasano1,Laget3}
\begin{eqnarray}
T^{off}_b &=& \frac{1}{32\pi^2 p}\int\limits_0^\infty \!\!dq \, u(q)
\frac{q}{E_q}\,
{  T}_M \, \, {  T}_A \nonumber \\
&\times& \ln \left|\frac{E_M{+}E_p{-}E_q{-}E_-}
{E_M{+}E_p{-}E_q{-}E_+}\right| ,
\label{off}
\end{eqnarray}
with 
\begin{eqnarray}
E_\pm^2 = (p \pm q)^2+m_{ex}^2,
\end{eqnarray}
where $m_{ex}$ is the mass of the exchanged meson.
In the present exploratory study we do not take into account
those off-shell contributions. We stress again that these
are highly model-dependent in any case.

Let us emphasize the general features of the reaction. It is clear that
the spectator proton momentum spectrum is sensitive to the reaction
mechanism. The nucleon exchange Born term $|T_a|^2$, given by
Eq.~(\ref{tree}), produces low and high momentum components, where the high
momentum part of the spectator distribution is dominated by $w(q)$, the 
$d$-wave component of the deuteron wave function. 
The $d$-wave component is more strongly model dependent and, therefore, the 
corresponding contribution of the Born term to the high momentum part of 
the spectator distribution is afflicted with some uncertainties.  
However, any really significant enhancement in that spectator distribution 
with respect to the results based on the Born term can  
be definitely attributed to meson rescattering processes on the 
spectator. Since the scattering amplitude $T_b$ is
directly proportional to the $Mp{\to}Mp$ amplitude $T_M$ and the annihilation
amplitude $T_A$ enters both the Born and rescattering diagrams, the 
enhancement in the high momentum part of the spectator distribution is
related to the relative magnitudes of their contributions and, thus,
essentially driven by the scattering amplitude $T_M$.

In any case, the isolation of the effects of $T_M$ from the data is by no
means a trivial task. There are also uncertainties related to possible contributions
from other processes besides those considered above, as well as uncertainties
related to the evaluation of off-shell corrections to the scattering amplitude
beyond Eq.~(\ref{off}). Nonetheless, the situation is still manageable since 
those uncertainties can be kept under control to some extent by selecting 
carefully the reaction kinematics -- see for example Ref.~\cite{Laget4}.

Our predictions for the ${\bar p}d{\to}D{\bar D}N$ reaction are
presented and thoroughly discussed in Section~\ref{sec:pd}.
In the Appendix we illustrate the applicability of the formalism to 
the proton spectator distributions as measured in the
${\bar p}d{\to}\pi^+2\pi^-{p}$ and ${\bar p}d{\to}2\pi^+3\pi^-{p}$ reactions.
Besides providing support for the reliability of the approach adopted for
the present investigation, those results are also useful for revealing 
similarities but also differences in the utilization of the reaction
$\bar p d \to nM N$ for exploring the $MN$ interaction for different
mesons and/or different kinematical regimes. 
 
All calculations are done with the deuteron wave function of the
CD-Bonn $NN$ potential~\cite{Machleidt}. But for shedding light on
the model dependence we also emply the wave functions of the
Paris \cite{Lacombe} and full Bonn \cite{MHE} $NN$ potentials.
We ignore the effects of
the Coulomb interaction in this exploratory study. Furthermore, 
we use averaged masses. Specifically, we use $m_D$ = 1866.9~MeV
for the mass of the $D$ and $\bar D$ mesons. 

\section{{The \boldmath$\bar{D}N$} and {\boldmath${D}N$} interactions}
\label{sec:interaction}

Considering charm and charge conservation antiproton-deuteron 
annihilation allows to study the following $D$-meson production reactions
\begin{eqnarray}
&&{\bar p}d \to D^0 {\bar D^0} n, \\
&&{\bar p}d \to D^+ {D^-} n, \\
&&{\bar p}d \to D^0 {D^-} p \ .
\end{eqnarray}
They involve the ${\bar D}N$ as well as the ${D}N$ scattering amplitude. 
Here we investigate only the ${\bar p}d{\to}D^0 D^-p$ reaction.
We select this channel because, in principle, all final particles can 
be measured for this reaction, which implies that there should be less 
uncertainties in the data evaluation discussed in the present study. 
For example, in case of the neutron as the spectator one needs to use the 
missing mass reconstruction technique. In this context let us also 
recall that charged open-charm mesons are in general reconstructed by the
leptonic, semileptonic and ha\-dron\-ic $K\pi\pi$ and $K\pi\pi\pi$ decay modes.
Usually the neutral open-charm mesons can be well detected by their hadron\-ic
$D^0{\to}K^-\pi^+$ and ${\bar D^0}{\to}K^+\pi^-$ decays~\cite{PDG}. For the full
reconstruction of the final state of the ${\bar p}d{\to}D^0D^-p$ reaction it is
therefore important that both semileptonic and hadronic modes are detected with
high accuracy. This should be kept in mind for the design of future
experiments, e.g., of the PANDA experiment~\cite{Panda}. High
accuracy is crucial for identification of $DN$ and ${\bar D}N$ rescattering
effects, whose absolute values and energy dependences could be very different, 
as indicated by several studies using different
models~\cite{Sibirtsev:1999js,Lin00,Hofmann05,Tolos04,Lutz06,Mizutani06}, 
and as will become clear from the present work, too. The identification of the 
different effects might be possible when a full reconstruction of the final 
state is feasible.
 
\subsection{The {\boldmath$\bar{D}N$} amplitude} 

\begin{figure}[t]
\vspace*{+1mm}
\centerline{\hspace*{4mm}\psfig{file=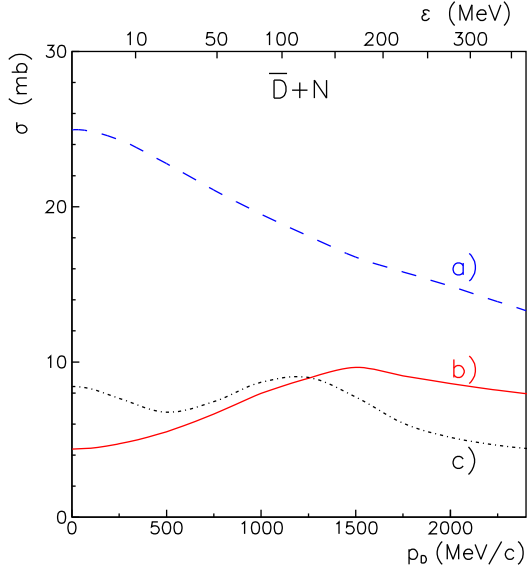,width=8.cm}}
\vspace*{-4mm}
\caption{Reaction cross section for (a) $D^-n{\to}D^-n$ (dashed line), 
(b) ${D^-}p{\to}{D^-}p$ (solid line) and (c) $D^-p{\to}{\bar D^0}n$ 
(dash-dotted line) as a function of the ${\bar D}$-meson momentum 
(lower axis) and the kinetic energy $\epsilon$ in the center-of-mass system 
(cms) (upper axis). 
}
\label{bras4}
\end{figure}

For the $\bar D N$ scattering amplitude we use the results of our recently 
published potential model~\cite{Haidenbauer}. This model for the $\bar D N$
interaction was constructed within the meson-exchange framework, but supplemented
with a short-distance contribution from one-gluon-exchange. The model was developed
in close analogy to the meson-exchange $KN$ interaction of the J\"ulich
group~\cite{Juel1,Juel2} utilizing SU(4) symmetry constraints.
The main ingredients of the interaction are provided by vector meson
($\rho$, $\omega$) exchange and higher-order box diagrams involving ${\bar D}^*N$,
$\bar D \Delta$, and ${\bar D}^*\Delta$ intermediate states. The short
range part is supplemented by additional contributions from genuine
quark-gluon processes~\cite{HHK,Juel3}. The reaction amplitude is 
obtained by solving a Lippmann-Schwinger type scattering equation for the 
interaction potential. 
The features of the $\bar D N$ amplitude based on this model are much 
more realistic than the ones employed in previous
studies~\cite{Sibirtsev:1999js,Cassing:1999wp,Sib01}. Indeed, in the former
studies the $D^-p$ cross section for instance was assumed to be momentum
independent and equal to ${\simeq}20$~mb \cite{Sibirtsev:1999js,Sib01} or
$5$~mb \cite{Cassing:1999wp}. Moreover, the angular dependence of the elastic
scattering was assumed to be either isotropic~\cite{Sibirtsev:1999js,Sib01} or
forward peaked~\cite{Cassing:1999wp}, i.e. proportional to $\exp(bt)$, where $t$ is the
four momentum transfer squared, with a slope $b{=}2$~GeV$^{-2}$. The reason for
such assumptions was the lack of any microscopic calculations of the ${\bar D}N$
scattering amplitude in those days.

To illustrate the differences between the previous
calculations~\cite{Sibirtsev:1999js,Cassing:1999wp,Sib01} and the
results of Ref.~\cite{Haidenbauer} we show in Fig.~\ref{bras4}
predictions for the $D^-n{\to}D^-n$,  ${D^-}p \to {D^-}p$
and $D^-p{\to}{\bar D^0}n$ reaction cross sections as a function
of the ${\bar D}$-meson momentum (lower axis) and the cm kinetic energy $\epsilon$
(upper axis). It is clear that the scattering cross sections for all channels 
depend significantly on the ${\bar D}$-meson momentum. 

\begin{figure}[b]
\vspace*{-10mm}
\centerline{\hspace*{4mm}\psfig{file=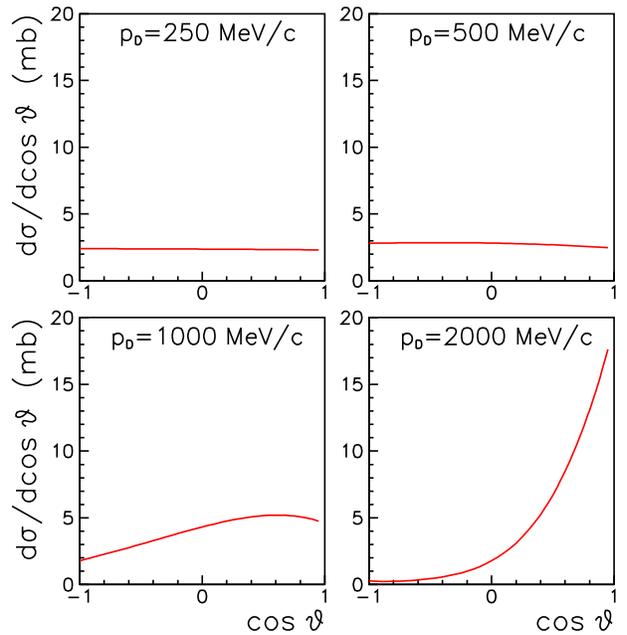,width=9.8cm}}
\vspace*{-6mm}
\caption{Differential cross sections for the $D^-p{\to}D^-p$ reaction
in the cm system at different momenta. }
\label{bras5}
\end{figure}

Note that the ${\bar D}N$ scattering amplitudes for the different reaction
channels shown in Fig.~\ref{bras4} are related to the isospin basis used 
in Ref.~\cite{Haidenbauer} by
\begin{eqnarray}
&&\hspace{-0.75cm}T_M (D^-n{\to}D^-n) = T_M ({\bar D^0} p{\to}{\bar D^0} p) 
= f_1 , \\
&&\hspace{-0.75cm}T_M (D^-p{\to}D^-p) = T_M ({\bar D^0} n{\to}{\bar D^0} n)
= \frac{1}{2}(f_0{+}f_1), \label{iso1} \\
&&\hspace{-0.75cm}T_M (D^-p{\to}{\bar D^0} n) = \frac{1}{2}(f_1 - f_0),
\end{eqnarray}
where $f_0$ and $f_1$ are the isospin $I{=}0$ and $I{=}1$ amplitudes,
respectively. 

Differential cross sections for $D^-p{\to}D^-p$ at different momenta 
are presented in Fig.~\ref{bras5}. The distributions are almost 
isotropic for momenta below $\simeq$500 MeV/c, but become forward peaked
at higher momenta. Note that there is no simple way to parametrize the angular
dependence with functions like $\exp(bt)$~\cite{Cassing:1999wp}, unless the 
slope parameter $b$ is taken to be momentum dependent.

For completeness, let us mention that other models of the $\bar DN$ interaction
have been published in recent years \cite{Hofmann05,Tolos08}. Those
authors considered $s$-waves only. The cross sections predicted by these 
models at threshold are 8.5~$mb$ ($D^-n \to D^-n$), 5.54~$mb$ 
($D^-p \to D^-p$), and 0.03~$mb$ ($D^-p \to \bar D^0n$) 
\cite{Hofmann05,Lutz06} and 10.6~$mb$, 2.64~$mb$, and 2.64~$mb$ 
for model~B of Ref.~\cite{Tolos08}, respectively.

\subsection{The {\boldmath$DN$} amplitude} 

The $DN$ interaction \cite{DN} employed in the present study is also constructed
in close analogy to the meson-exchange $\bar K N$ model of the J\"ulich group
\cite{MG} as well as by exploiting the close connection between the $\bar DN$
and $DN$ systems due to G-parity conservation. Specifically, the latter
constraint fixes the contributions to the direct $DN$ interaction
potential while the former one provides the transitions to and interactions in
channels that can couple to the $DN$ system. 
Accordingly, the $DN$ interaction is likewise provided by vector-meson ($\rho$,
$\omega$) exchange and higher-order box diagrams involving ${D}^*N$, $D \Delta$,
and ${D}^*\Delta$ intermediate states. The short-ranged quark-gluon processes,
however, are absent here because the quark-exchange mechanism cannot contribute
to the $DN$ interaction due to the different quark structure of the $D$ meson. 
As far as the coupling to other channels is concerned, we follow here the
arguments of Ref.~\cite{MG} and we take into account only the channels
$\pi\Lambda_c(2285)$ and $\pi\Sigma_c(2455)$. Furthermore, we restrict ourselves
to vector-meson exchange and we do not consider any higher-order diagrams in
those channels. Pole diagrams due to the $\Lambda_c(2285)$ and $\Sigma_c(2455)$
intermediate states are, however, consistently included in all channels. 

\begin{figure}[b]
\vspace*{-5mm}
\centerline{\hspace*{4mm}\psfig{file=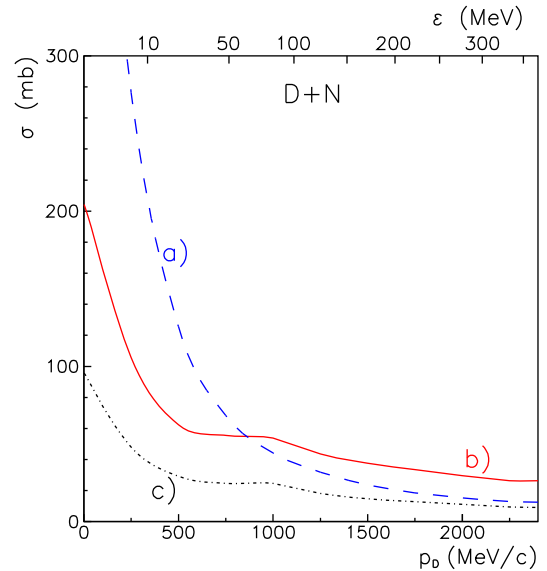,width=8.cm}}
\vspace*{-4mm}
\caption{Reaction cross sections for (a) $D^0n{\to}D^0n$ (dashed line),  
(b) $D^0p{\to}D^0p$ (solid line) and (c) $D^0p{\to}D^+n$ (dash-dotted line) 
as a function of the ${\bar D}$-meson momentum (lower axis) and the cms
kinetic energy $\epsilon$ (upper axis). }
\label{bras4c}
\end{figure}

In this basic model all free parameters - the coupling constants and the cut-off
masses at the vertex form factors of the occurring meson-meson-meson and
meson-baryon-baryon vertices, cf. \cite{MG} - are fixed by the assumed SU(4)
symmetry and the connection with the $\bar KN$ model, respectively. When solving
the coupled-channel Lippmann-Schwinger equation with this interaction model we
observe that two states are generated dynamically below the $DN$ threshold, one
in the $S_{01}$ partial wave and the other one in the $S_{11}$ partial wave. 
(We use here the standard spectroscopical nomenclature $L_{I \, 2J}$.) In view
of the close analogy between our $DN$ model and the corresponding $\bar KN$
interaction \cite{MG} this is not too surprising, because also the latter yields
a quasi-bound state in the $S_{01}$ channel which is associated with the
$\Lambda (1405)$ resonance. The bound states in both the $\bar KN$ and $DN$ are
generated by the strongly attractive interaction due to the combined effect of
$\omega$, $\rho$ and scalar-meson exchanges, which add up coherently in specific
channels. 

It should be said that studies of the $\bar KN$ and $DN$ interaction within
chiral unitary (and related) approaches likewise generate the $\Lambda (1405)$
resonance dynamically but also states in the $DN$ system
\cite{Hofmann05,Mizutani06,Lutz04}. In those approaches the strong attraction is
also provided by vector-meson exchange \cite{Hofmann05} or by the
Weinberg-Tomazawa term \cite{Mizutani06,Lutz04}. In
Refs.~\cite{Tolos04,Mizutani06,Tolos08} the authors argued that the state
occuring in the $S_{01}$ channel of the charm $C = 1$ sector should be identified 
with the $I=0$ resonance $\Lambda_c(2593)$. We adopt this viewpoint here too. 
Furthermore, we identify the state we get in the $S_{11}$ channel with the
$I=1$ resonance $\Sigma_c(2800)$ \cite{PDG}. 

In order to make sure that the $DN$ model we are going to apply in our study of
the reaction ${\bar p}d{\to}D^0 {D^-}p$ incorporates these features also
quantitatively we fine-tune the contributions of the scalar mesons to the $DN$
interaction so that the position of those states generated by the model coincide
with the values given in the list of the Particle Data Group. This can be
achieved by a moderate change in the coupling constants of the $\sigma$ meson
(from $1$ to $2.6$) and the $a_0$ meson (from $-2.6$ to $-4.6$), cf. Table 2 in
Ref.~\cite{Haidenbauer}. 

\begin{figure}[t]
\vspace*{-6mm}
\centerline{\hspace*{4mm}\psfig{file=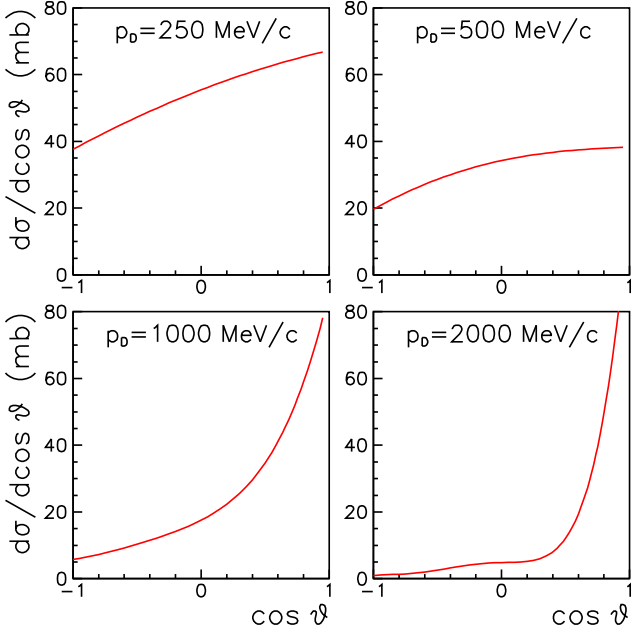,width=9.8cm}}
\vspace*{-6mm}
\caption{Differential cross sections for the $D^0p{\to}D^0p$ reaction
in the cm system at different momenta. }
\label{bras5a}
\end{figure}
 
Interestingly, our model generates a further state, name\-ly in the $P_{01}$
partial wave, which, after the above fine-tuning, lies at 2803 MeV, i.e. just
below the $DN$ threshold. We are tempted to identfy this state with the
$\Lambda_c(2765)$ resonance, whose quantum numbers are not yet established
\cite{PDG}. Though we do not reproduce the resonance energy quantitatively, we
believe that further refinements in the $DN$ model, specifically the inclusion
of the $\Lambda_c\pi\pi$ channel in terms of an effective $\sigma\Lambda_c$
channel, can provide sufficient additional attraction for obtaining also
quantitative agreement. The mechanism could be the same as in case of the Roper
($N^*(1440)$) resonance, which is generated dynamically in the J\"ulich $\pi N$
model \cite{Krehl,Achot}. Here the required strong attraction is produced via
the coupling of the $\pi N$ $p$-wave (where the Roper occurs) to the $s$-wave in
the $\sigma N$ system, facilitated by the different parities of the $\pi$ and
$\sigma$ mesons.

Some results of our $DN$ model are presented in Figs.~\ref{bras4c} and
\ref{bras5a}. The $DN$ scattering amplitude for the different reaction channels
shown in the Fig.~\ref{bras4c} are related to those in the isospin basis by 
\begin{eqnarray}
&&\hspace{-0.75cm}T_M (D^0n{\to}D^0n) = T_M ({D^+} p{\to}{D^+} p) = f_1 , \\
&&\hspace{-0.75cm}T_M (D^0p{\to}D^0p) = T_M ({D^+} n{\to}{D^+} n)
= \frac{1}{2}(f_0{+}f_1), \label{iso1a} \\
&&\hspace{-0.75cm}T_M (D^0p{\to}{\bar D^+} n) = \frac{1}{2}(f_1 - f_0),
\end{eqnarray}
where $f_0$ and $f_1$ are the isospin $I{=}0$ and $I{=}1$ amplitudes
respectively. 

Obviously, also the $DN$ cross sections show a significant momentum dependence
in all charge channels. Furthermore, the cross sections are substantially larger
than those we obtain for $\bar DN$. Specifically, for the pure $I=1$ channel 
$D^0n$ the cross section amounts to almost 600 mb at threshold. This is not too
surprising in view of the near-by quasi-bound state. The latter is also
reflected in the $s$-wave scattering lengths, 
\begin{eqnarray}
\nonumber
&&a^{I=0}_{DN} = (-0.41 + {\it i} 0.04) \ {\rm fm} \\
&&a^{I=1}_{DN} = (-2.07 + {\it i} 0.57) \ {\rm fm} \ ,
\end{eqnarray}
namely by the rather large value of the real part in the $I=1$ channel. For
completeness, let us mention here that the scattering lengths of the $DN$
interaction of Hofmann and Lutz \cite{Hofmann05}, reported in \cite{Lutz06},
amount to about $-0.4$~fm for both isospin channels. In agreement with that work we 
find that the imaginary part is negligibly small for $I=0$. However, contrary to
\cite{Lutz06} in our $DN$ model this is not the case for the $I=1$ channel.

Angular distributions for the reaction $D^0p{\to}D^0p$ are shown in
Fig.~\ref{bras5a}. Obviously, there is a strong anisotropy already at fairly low
momenta. It is due to significant contributions in the $P_{01}$ partial wave in
this momentum region induced by the near-threshold quasi-bound state produced by
our model, as discussed above. For higher momenta the differential cross section
becomes forward peaked, similar to the predictions of our model for the $\bar DN$
system. 

Further details of our $DN$ model will be reported in a forthcoming publication
\cite{DN}. 
%%%%%%%%%%% DN Interaction end %%%%%%%%%%%%%%%%%

\section{ {\boldmath$\bar{D}D$} production in {\boldmath$\bar{p}N$}
annihilation} 
\label{sec:annihilation}

\begin{figure}[t]
\vspace*{-5mm}
\centerline{\hspace*{4mm}\psfig{file=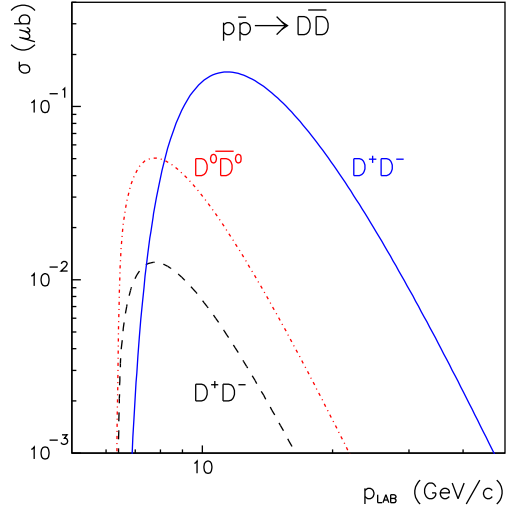,width=8.cm}}
\vspace*{-4mm}
\caption{Predictions for the $\bar pp\to \bar DD$
annhilation cross section taken from Refs.~\cite{Kroll89} (solid line)
and \cite{Kai94} (dashed and dash-dotted lines). 
}
\label{bras20a}
\end{figure}

The total cross section for the reaction $\bar{p}d \to D^0D^-p$ depends 
crucially on the elementary $\bar NN \to \bar DD$ annihilation amplitude $T_A$. 
Unfortunately, so far there is no experimental information about this
reaction and even theoretical studies are rather scarce 
\cite{Kroll89,Kai94,Kerbi95}. In Ref.~\cite{Kroll89} results were given
for the reaction $\bar pp \to D^-D^+$ in a quark plus diquark model, 
where the elementary flavour changing process is due to 
diquark-antidiquark annihilation and subsequent creation of a 
quark-antiquark pair through one gluon. 
Kaidalov and Volkovitsky calculated the reaction $\bar pp \to \bar DD$
in the framework of a non-perturbative quark-gluon string model, based on
secondary Regge pole exchanges including absorptive corrections~\cite{Kai94}.
The result of both works are summarized in Fig.~\ref{bras20a}.
The larger cross sections are predicted by the model of Ref.~\cite{Kroll89}
with a maximal value of around 0.15~$\mu b$ 
at $p_{lab} =$~12 GeV/c corresponding to $\sqrt{s} \approx 5$~GeV. The
model of Ref.~\cite{Kai94} yields $\sigma_{\bar pp \to \bar D^0D^0} 
\approx 0.05 \ \mu b$ at the maximum, while the one for $\bar pp \to \bar D^-D^+$
is a factor of 4 smaller. In both approaches, the unknown parameters of
the models were fixed by considering corresponding flavour changing
reactions involving strange quarks 
($\bar pp \to \bar KK$ and/or $\bar pp \to \bar\Lambda \Lambda$, etc.). 

Superficially the predictions of the two studies are qualitatively 
similar, specifically for ${\bar pp \to \bar D^0D^0}$ which in case of the
model of Kroll et al. \cite{Kroll89} is suppressed by roughly a factor four 
as compared to the prediction for $\bar pp \to D^-D^+$ \cite{Schweiger}
so that its maximal value would then practically coincide with the one 
obtained in \cite{Kai94}. However, for the reaction $\bar pp \to D^-D^+$
the magnitudes of the cross sections differ by more than a factor of 10. 
Furthermore, there is a significant difference in the energy dependence of the 
two model predictions, as seen in Fig.~\ref{bras20a}, so that the variations 
of the predictions are even larger in specific energy regions. 

Admittedly, those results are only of limited use for our own 
investigation of the $\bar p d\to D^-D^0 p$ reaction. 
For example, for the amplitude corresponding
to the reaction chain $\bar p d \to D^-D^0 p \to D^-D^0 p$, involving
rescattering in the $D^0 p \to D^0 p$ and $D^- p \to D^- p$ channels, 
one needs the pure $I=1$ annihilation amplitude $\bar p n \to D^-D^0$
which cannot be reconstructed from the available information about
those models. The other contributions to the reaction amplitude,
involving ($DN$ or $\bar DN$) charge-exchange rescattering,
$\bar p d \to D^-D^+ n \to D^-D^0 p$ and 
$\bar p d \to \bar D^0D^0 n \to D^-D^0 p$,
require $\bar pp \to D^-D^+$ and $\bar pp \to \bar D^0D^0$,
respectively, but here the relative phase between the 
terms is not known. 
Thus, we are facing the problem that we either have to add all
contributions incoherently and make additional assumptions about
the isospin dependence of $\bar NN \to \bar DD$ or we consider only 
the amplitude involving elastic $DN$ and $\bar DN$ rescattering.
We prefer the latter option. In this case we can add the Born term and
the $DN$ and $\bar DN$ rescattering contributions coherently, because
they all involve the same elementary $\bar p n \to D^-D^0$ annihilation
amplitude, and we can include the resulting interference effects in 
the evaluation of the observables. However, absolute predictions are 
out of reach and all results will be shown as number of events only. 
On the other hand we consider the energy dependence of the elementary 
$\bar p n \to D^-D^0$ annihilation amplitude in our calculation by 
adopting the results given in \cite{Kai94} for $\bar p p \to \bar DD$.
But we should say that its influence on the observables shown in the
present paper is practically negligible. 

\section{Open charm production in {\boldmath$\bar{p}d$} annihilation}
\label{sec:pd}

\begin{figure}[b]
\vspace*{-5mm}
\centerline{\hspace*{3mm}
\psfig{file=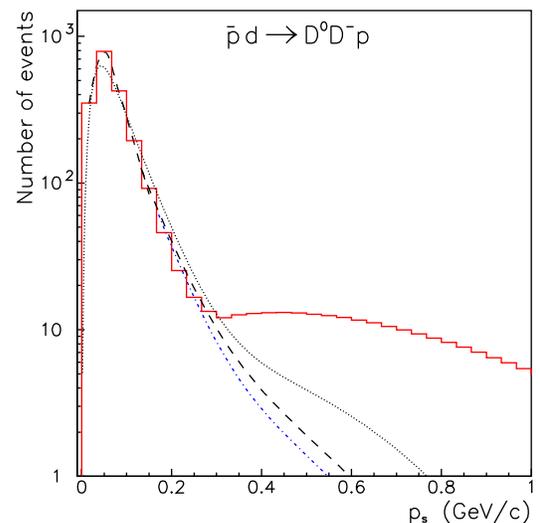,width=8.cm}}
\vspace*{-5mm}
\caption{Proton momentum spectrum for the
${\bar p}d{\to}D^0 {D^-}p$ reaction. The dashed (dotted, dash-dotted) line 
shows the result of a calculation for the nucleon exchange Born diagram only 
(Eq.~(\ref{tree})) based on the $s$ and $d$-wave parts 
of the deuteron wave function of the CD Bonn \cite{Machleidt} 
(Paris \cite{Lacombe}, full Bonn \cite{MHE}) $NN$ potential. 
The solid histogram is the full calculation (for CD Bonn) that includes $D^0p$ and 
${D^-}p$ rescattering.
}
\label{bras3}
\end{figure}
Now we present results for $D^0D^-$ production in antiproton-deuteron 
annihilation utilizing the formalism and the elementary $DN$ and $\bar DN$ 
amplitudes described above, taking into account the Born diagram of
Fig.~\ref{diag}a) and the rescattering diagram of Fig.~\ref{diag}b).
For the 
latter we consider both $D^-p$ and $D^0p$ scattering in the final state. 
The ${\bar p}n{\to}D{\bar D}$ threshold on a free nucleon corresponds to the
antiproton momentum of roughly $6.43$~GeV/c. The absolute threshold for the
${\bar D}D$ production in antiproton-deuteron annihilation is at the
antiproton momentum of $4.55$~GeV/c. Evidently, close to the reaction
threshold the production rate will be strongly suppressed by the phase space. 
We choose for our calculation the antiproton momentum of $7$~GeV/c, 
corresponding to the region where the model calculation of \cite{Kai94} 
predicts the largest cross sections for the elementary $\bar NN \to \bar DD$ 
reaction. 

\begin{figure*}[t]
\vspace*{-5mm}
\centerline{\hspace*{3mm}\psfig{file=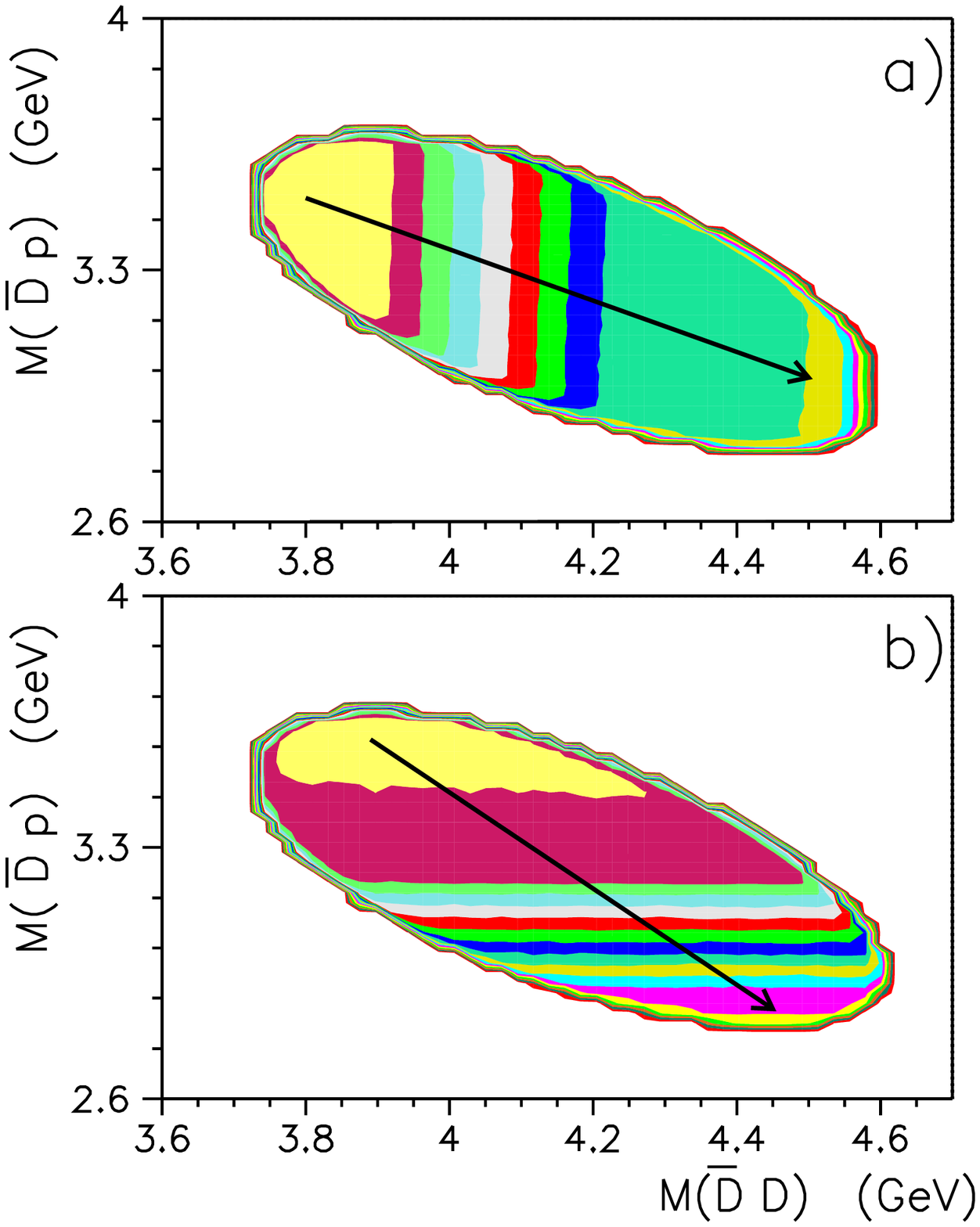,width=7.cm}\hspace*{-4mm}
\psfig{file=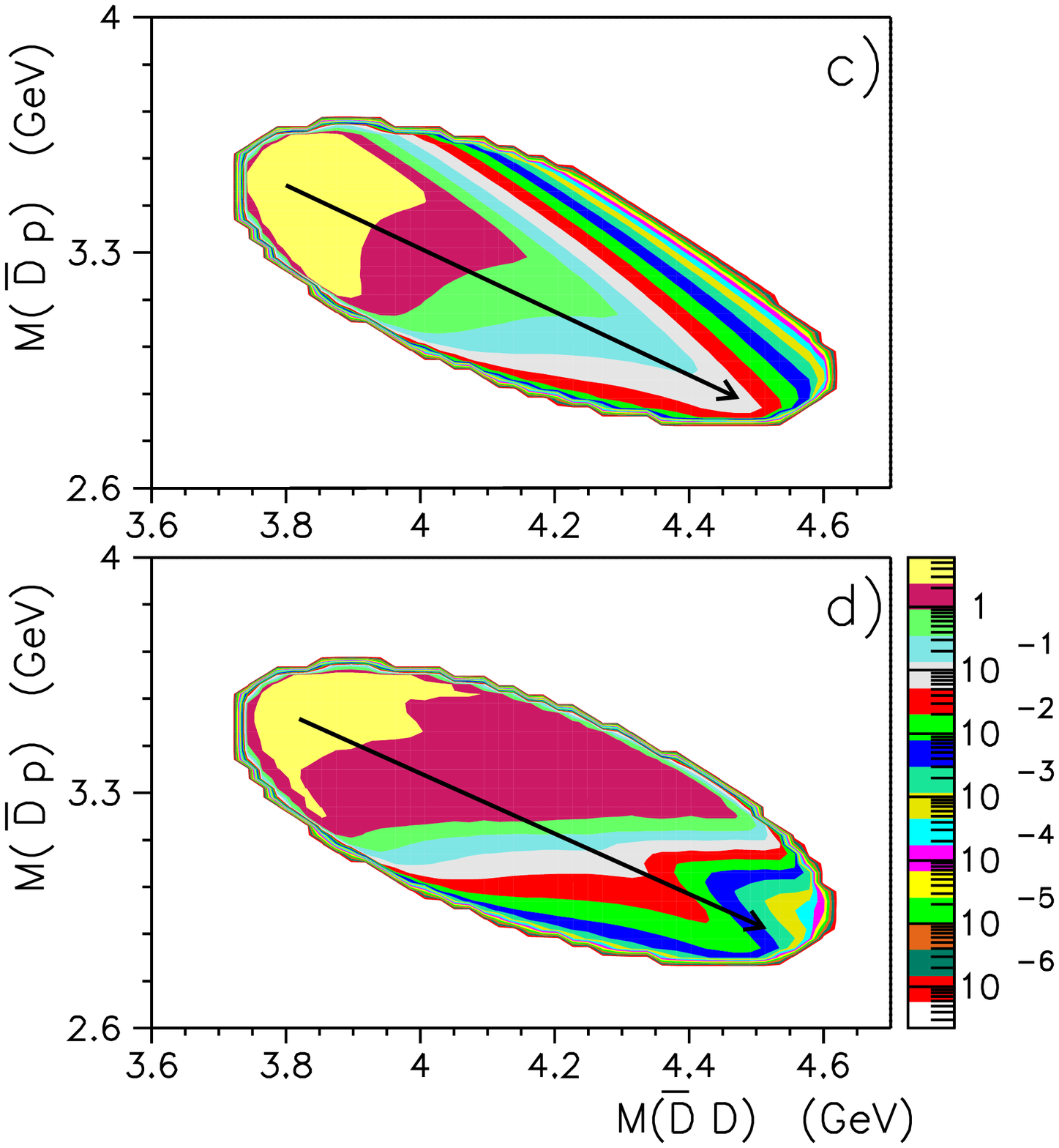,width=7.cm }}
\vspace*{-3mm}
\caption{Dalitz plot for the ${\bar p}d{\to}D^0 {D^-}p$ reaction. Here, 
$M({\bar D}p)$ and $M({\bar D}D)$ are the ${D^-}p$ and $D^0D^-$ invariant
masses, respectively. The arrows points in the direction of decreasing 
intensity of the distribution. Each plot a)-d) is explained in the text.
}
\label{bras6}
\end{figure*}

\subsection{Spectator momentum distribution}

In Fig.~\ref{bras3} we present our predictions for the spectator
proton momentum distribution. Here the solid histogram indicates 
the full result that includes the Born and the rescattering 
diagrams while the dashed line is the result based on the nucleon-exchange 
Born diagram alone, both obtained with the deuteron wave function of
the CD Bonn $NN$ potential \cite{Machleidt}. Since the absolute
normalization of the reaction cross section is quite uncertain we show the
results as number of generated events. 

The dotted and dash-dotted curves are results for the Born term alone employing
the deuteron wave functions of the Paris \cite{Lacombe} and full Bonn \cite{MHE}
potentials, respectively. Obviously, there is some model dependence which 
becomes more pronounced for spectator proton momenta above 300 MeV/c. 
But the rescattering mechanism is definitely by far the most dominant 
effect for momenta from around 400 MeV/c upwards. 
Since rescattering occurs in both $D^-p$ and $D^0p$ systems one needs 
to apply specific methods to separate their contributions in a reliable way, 
as will be discussed below.

In comparison to the multipion production case, cf. Figs.~\ref{bras1}
and \ref{bras2} in the Appendix, the enhancement due to the rescattering
processes sets in at noticably higher spectator momenta and is also 
less pronounced. It was argued \cite{Locher0,Kolybasov1} that the strong
enhancement seen in the proton momentum spectrum for the multipion
reactions is to a good part due to the excitation of the $\Delta (1232)$
resonance in the $\pi N$ rescattering processes. Though our $DN$
scattering amplitude is dominated likewise by poles, in several partial
waves, cf. the discussion above, their influence on the momentum spectrum
seems to be smaller, presumably because they all lie below the $DN$ elastic
threshold. 
The enhancement we get for the ${\bar p}d{\to}D^0 {D^-}p$ reaction seems
to be somewhat smaller than what was reported in an earlier model
calculation by Cassing et al.~\cite{Cassing:1999wp}. But one has to keep
in mind that in the latter work the contribution of the $d$-wave component
to the Born (spectator) term was neglected and, moreover, the results for
the Born term and the rescattering term are shown separately, while we
added them coherently in our calculation. 

\subsection{Dalitz plot}

A well-known method for the reconstruction of the reaction dynamics
is the Dalitz plot analysis of the final
state~\cite{Byckling,Hanhart,SibirtsevK,SibirtsevK1,TOF}. 
For instance, for multi-pion production from $\bar p d$ annihilation
the presence of rescattering effects was demonstrated via the
projection of the Dalitz plot on the invariant mass spectrum of the
final $\pi p$ system~\cite{Voronov,SibirtsevAF}. 
An analysis in form of a partial wave decomposition of the
Dalitz plot was proposed~\cite{Hanhart,SibirtsevK,SibirtsevK1}
for the $pp{\to}pK^+\Lambda$ reaction, which finally allowed to study and 
separate~\cite{TOF} non-resonant and resonant contributions in the 
$K^+\Lambda$ subsystem.
A similar technique could be applied in the analysis of the ${\bar p}d{\to}D^0
{D^-}p$ reaction. This method allows to study all subsystems, ${D^0}p$, ${D^-}p$ 
and $D^0D^-$, but obviously requires high mass resolution and significant
statistics. 

Fig.~\ref{bras6} presents the Dalitz plot evaluated for the reaction ${\bar p}d{\to}D^0
{D^-}p$ at the antiproton momentum of $7$~GeV/c. The horizontal axes in the
plot indicate the invariant mass of the $D^0D^-$ system, while the vertical axes
indicate the mass of the ${D^-}p$ system. Here panel a) shows the results
obtained with the nucleon-exchange Born diagram only, b) those 
obtained with $D^-p$ rescattering alone, while c) illustrates the
results with $D^0p$ FSI alone. Finally, Fig.~\ref{bras6}d) contains 
the full results, {\it i.e.} when all three contributions are included coherently.
The arrows in the figures indicate the direction of decreasing intensity 
of the distribution. Note that we implemented a cut on the spectator proton 
momentum in the calculations in order to reduce the contribution from the Born 
diagram. Specifically, we considered only events involving spectator protons
with momenta above $300$~MeV/c.

Evidently, the differences between the distributions resulting from the
different diagrams are quite significant. The result based on the Born term alone
indicates strong correlations between the $D^0$- and $D^-$-meson invariant mass.
This is due to the fact that they are produced from the same vertex,
namely via ${\bar p}n{\to}D^0D^-$. The invariant energy of the $D^0D^-$ system
is essentially given by the energy of the incoming antiproton, while the dispersion
of the distribution is related to the square of the deuteron wave function. Assuming
the target neutron to be at rest, the invariant mass of the $D^0D^-$ system
produced in the  reaction ${\bar p}n{\to}D^0D^-$ at antiproton momentum of $7$~GeV/c is
equal to $3.86$~GeV. The strong $D^0D^-$ correlation produces also 
a kinematical reflection, detectable in the $D^-p$ system, namely in form of an 
enhancement in the high mass $D^-p$ spectrum. 
However, this enhancement, being purely kinematical, does not have any 
relevance for the interpretation of the rescattering mechanism.

Results obtained for the rescattering diagrams alone are shown in
 Figs.~\ref{bras6}b) and c). It is clear that there are no strong correlations
in the $D^0D^-$ system anymore. Now the spectrum is primarily distorted by the 
corresponding rescattering terms. The ${D^-}p$ and $D^0p$ projections of the 
Dalitz plot show the distribution produced by the relevant scattering amplitudes.

The final distribution, shown in Fig.~\ref{bras6}d), corresponds to a 
calculation that includes the Born diagram plus both rescattering diagrams. 
It is fairly non-uniform and clearly indicates substructures resulting from 
the individual reaction mechanisms. 

\subsection{Missing mass of the exchanged meson}
A method~\cite{Barmin} that could be useful for the separation of the $\bar D N$ 
and $DN$ rescattering contributions is 
based on the assumption that the dominant part of the rescattering amplitude 
comes from contributions where the particles in the intermediate state
are on shell. 
We should emphasize that this method has no physical meaning when
it comes to the off-shell part of the rescattering diagrams because then 
the exchanged meson is virtual.

Under the assumption that the charmed meson and nucleon are on-shell
before undergoing rescattering, one can reconstruct the four-momentum of 
the meson in the loop following the missing mass technique via 
\begin{figure}[t]
\vspace*{-5mm}
\centerline{\hspace*{3mm}
\psfig{file=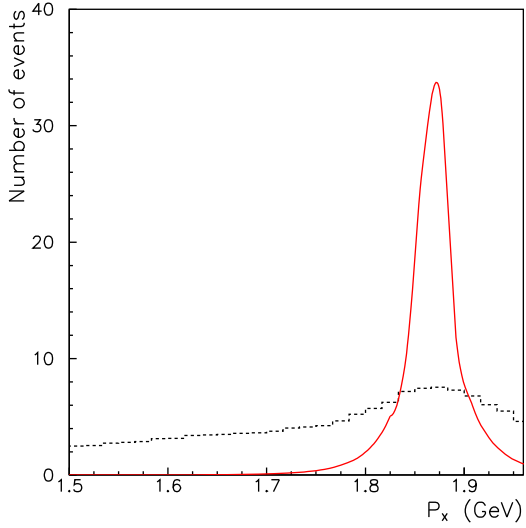,width=8cm}}
\vspace*{-4mm}
\caption{The missing mass distribution of $P_X$ given by 
Eq.~(\ref{recon2}) for the ${\bar p}d{\to}D^0 {D^-}p$ reaction. 
The dashed histogram shows results obtained with the nucleon-exchange 
Born diagram and the $D^0p$ FSI for spectator proton momenta above
300~MeV/c.
The solid histogram is the result for the $D^-p$ rescattering diagram 
based on Eq.~(\ref{rescat}).}
\label{bras7}
\end{figure}

\begin{eqnarray}
P_X^2 = (P_s+P_{D^-}-P_N)^2,
\label{recon2}
\end{eqnarray}
where $P_s$ and $P_{D^-}$ are the four-momenta of the spectator proton and
the final $D^-$ meson, and $P_N$ is the four-momentum of the nucleon 
in the loop (i.e. the one involved in the rescattering process), 
given by the loop momentum $q$ and energy $E_q$ 
as in Eq.~(\ref{rescat}). 
Let us assume for the moment that the scattering process takes place on
a free nucleon at rest. Then we would get $P_X^2{=}m_D^2$, where $m_D$ 
is the mass of the (incoming) $D$ meson. 
However, since the reaction does not take place on a free nucleon but
on a nucleon from the initial deuteron, the interacting nucleon is not 
at rest and, therefore, one expects a distribution of the 
missing mass in Eq.~(\ref{recon2}) around the central value of $m_D$  
reflecting the Fermi motion of the nucleon in the deuteron.

It is clear that the Born term as well as $D^0p$ scattering would not
lead to such a distribution because in this case there is no correlation 
between the four-momenta in Eq.~(\ref{recon2}). 

In Fig.~\ref{bras7} we present results for the missing mass of the exchanged 
meson as given by Eq.~(\ref{recon2}), obtained for the antiproton momentum of 
$7$~GeV/c. In the corresponding calculations the Born diagram and 
the rescattering amplitudes according to Eq.~(\ref{rescat}) are taken
into account. Then we evaluate Eq.~(\ref{recon2}) using the four-momenta of 
the final spectator proton and of the final $D^-$-meson,
and assume that $P_N{=}(m_p,{\bf 0})$. 
The dashed histogram in Fig.~\ref{bras7} includes contributions from the 
nucleon-exchange Born diagram for spectator momenta above $300$~MeV/c as well
as from the $D^0p$ FSI. The solid histogram is the result obtained for the 
$D^-p$ rescattering term. 

The results shown in the Fig.~\ref{bras7} look very promising with regard to
the possibility for a separation of the reaction mechanisms. 
However, one should keep in mind that there are uncertainties due to the
on-shell assumption made in the model calculation as well as in the evaluation of
the missing mass, which cannot be quantified easily. 
Nonetheless, we want to mention that 
this method was actually used in Ref.~\cite{Barmin} in the data evaluation and
reconstruction of the hyperon production mechanisms in antiproton annihilation on
xenon nuclei. Recently, this missing-mass method was also utilized for the analysis
of $K^+K^-$ pair production from carbon~\cite{ANKE}. It is important to stress
that the method cannot and should not be applied for too high momenta of the 
spectator proton, where the reaction is dominated by off-shell contributions 
and, therefore, the basic assumptions of the method 
are evidently no longer valid. 

We have also performed calculations of the missing mass distribution  
for the corresponding case of a final $D^0$ meson where one can isolate the 
$D^0p$ rescattering contributions. The results are qualitatively very
similar to the $D^-$ case and therefore we do not show them here. 

\subsection{Correlation between the scattering planes }

Finally, we discuss the correlation between the two scattering
planes~\cite{Golubeva}. One plane is given by the momenta of the antiproton and
of the spectator proton. The other one is fixed by the momenta of the antiproton
and of the produced charmed meson, the $D^-$ meson, say. Then, due to the 
conservation of the transverse momenta in the
$D^-p{\to}D^-p$ scattering process the azimuthal angle between these planes is 
peaked
around $\phi{\simeq}180^\circ$. This correlation simply follows from the reaction
kinematics. If the spectator proton is at rest before the scattering and the
$D^-$-meson has no transverse momentum, then, after $D^-p{\to}D^-p$ scattering
the transverse components of the momenta of the final $D^-$-meson and proton 
must be exactly the same but aligned in opposite direction.  
However, both the Fermi motion in the deuteron and the ${\bar
p}n{\to}D^-D^0$ annihilation allow for some variations in the transverse 
momenta of the spectator proton and of the $D^-$-meson. That is why in an 
actual experiment one would expect a distribution of the azimuthal angle
around the value $\phi{=}180^\circ$.

\begin{figure}[t]
\vspace*{-5mm}
\centerline{\hspace*{3mm}
\psfig{file=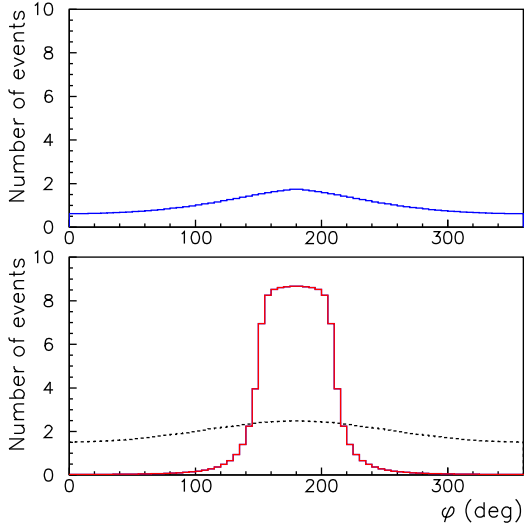,width=8cm}}
\vspace*{-4mm}
\caption{The distribution of the azimuthal angle between the scattering plane
given by the momenta of the antiproton and spectator proton and the plane 
fixed by the momenta of the antiproton and the {\bf $D^-$}-meson. 
The results are for spectator protons with momenta above $300$~MeV/c. 
The upper panel shows the result for the Born diagram alone, while in the 
lower panel calculations including the $D^-p$ and $D^0p$ rescattering 
diagrams are presented. 
Here the solid histogram shows the distribution obtained for 
rescattering of the $D^-$-meson, while the 
dashed histogram is the corresponding result for 
rescattering of the $D^0$-meson. 
}
\label{bras8}
\end{figure}

Results of our model calculation for the distribution of the azimuthal angle 
are presented in Fig.~\ref{bras8}. They are obtained again by imposing a cut
on the spectator proton momentum so that only momenta above $300$~MeV/c 
contribute. The scattering plane is fixed by the momenta of the antiproton
and the $D^-$-meson. 
The upper panel of Fig.~\ref{bras8} shows predictions for the 
nucleon-exchange Born diagram while the lower panel contains results 
including the rescattering diagrams. 
Here the solid histogram corresponds to the contributions of the $D^-p$ 
rescattering diagram and the dashed histogram to those from 
$D^0p$ rescattering. From these results it seems feasible that the 
two rescattering contributions can be well isolated.

Also here we have considered the corresponding case for the $D^0$ meson,
where again the results turned out to be qualitatively similar. 

\section{Summary}
\label{sec:summary}

In this paper we 
examined the possibility to extract information about the $DN$ and ${\bar D}N$
interactions from the ${\bar p}d{\to}D^0 {D^-}p$ reaction.
We utilized the notion that those open-charm mesons are first produced by 
annihilating antiprotons on the deuteron and subsequently rescatter on the 
remaining (spectator) nucleon. The latter process is then exploited for 
investigating the $DN$ and ${\bar D}N$ interactions.
To explore the potential of a corresponding experiment we performed concrete 
model calculations taking into account the nucleon-exchange Born diagram 
as well as rescattering diagrams. 

As a test of the approach we first applied it to multipion production in 
${\bar p}d$ annihilation and we compared our results with data available 
for the ${\bar p}d{\to}\pi^+2\pi^-{p}$ and
${\bar p}d{\to}2\pi^+3\pi^-{p}$ reactions. These
data~\cite{Riedlberger,Ahmad,Zemany} show strong evidence for 
contributions from $\pi{N}{\to}\pi{N}$ rescattering, which can be seen in 
the spectra of the spectator proton and invariant mass distribution 
of the final pion and proton.  We obtained very reasonable agreement with 
the data available for the spectator proton distributions.

In our investigation 
of the ${\bar p}d{\to}D^0 {D^-}p$ reaction we utilized realistic
${\bar D}N{\to}{\bar D}N$~\cite{Haidenbauer} and $DN{\to}DN$~\cite{DN} 
scattering amplitudes. This is a substantial improvement over 
previous studies~\cite{Sibirtsev:1999js,Cassing:1999wp,Sib01} which 
employed simplistic ${\bar D}N$ and $DN$ scattering amplitudes based on
somewhat questionable assumptions.
 
We found that below spectator momenta of around 300~MeV/c 
the reaction is dominated by the nucleon-exchange Born diagram.
For higher spectator momenta there is a sizable contribution 
from the rescattering diagrams. In particular, their contribution is 
significantly larger than the uncertainties due variations in the high-momentum 
component of the deuteron wave function. Thus, selecting events with 
spectator momenta above 300 or 400~MeV/c, say, should allow to obtain a 
data sample that can be used for extracting information about the 
$DN$ and $\bar DN$ interactions. 

Subsequently 
we explored different methods for isolating the contributions from 
the $DN$ and $\bar DN$ rescattering terms. We showed that the missing 
mass technique and the
correlation between the planes given by the scattered meson and nucleon allow
a reasonable reconstruction of the reaction dynamics and to separate the 
contributions of ${\bar D}N$ rescattering from those of $DN$ rescattering. 
Since these methods are based on the reaction kinematics we consider them 
as promising tools to extract information on the ${\bar D}N$ and $DN$ 
interactions from the reaction  ${\bar p}d{\to}D^0 {D^-}p$. 

\begin{acknowledgement}
We appreciate discussions with
A.~Afanasev, W.~Melnitchouk, W.~Schweiger, S.~Stepanyan, K.~Tsushima 
and B.~Wojtsekhowski.
This work was financially supported by the
Deutsche For\-schungs\-gemeinschaft (Project no. 444 BRA-113/14 
and through funds provided by the SFB/TR 16
``Subnuclear Structure of Matter'')
and the Brazilian agencies CAPES, CNPq and FAPESP.
It was also supported in part by the Helmholtz Association
through funds provided to the virtual institute ``Spin and strong
QCD'' (VH-VI-231).
This research is part of the EU
Integrated Infrastructure Initiative Hadron Physics Project under
contract  number RII3-CT-2004-506078. A.S. acknowledges support by
the JLab grant SURA-06-C0452 and the COSY FFE grant No. 41760632 
(COSY-085). 
\end{acknowledgement}

\appendix

\section{Multi-pion production in {\boldmath$\bar{p}d$} annihilation}
\label{sec:multpi}
To illustrate the applicability of the discussed formalism we consider
experimental data available for the reactions ${\bar p}d{\to}2\pi^+3\pi^-{p}$
and ${\bar p}d{\to}\pi^+2\pi^-{p}$ obtained at the Low Energy Antiproton Ring
(LEAR) at CERN using annihilation at rest in a hydrogen
gas~\cite{Riedlberger,Ahmad} and at the Brookhaven National Laboratory (BNL)
using the deuterium bubble chamber~\cite{Zemany}. 

We calculate the proton
spectator distribution by summing the Born and rescattering amplitudes and
integrating over the 6-body and 4-body final states. 
The calculations are done with the deuteron wave function of the CD-Bonn
potential~\cite{Machleidt}. Available data~\cite{Vandermeulen1,Amsler} on pion
multiplicities for $\bar pp$ annihilation at rest and in flight show 
practically no dependence on the antiproton momentum within the range up 
to $\simeq$100~MeV/c. Therefore, we assume the annihilation amplitude 
$T_A$ to be a constant. The spectator proton momentum distribution was
measured in different experiments~\cite{Riedlberger,Ahmad,Zemany} 
and the data were published with arbitrary normalization depending on the total 
number of detected events. Therefore, we normalize our calculation to the data
but we also normalize the different data sets to each other, as explained 
below. 

For evaluating the contribution from pion-nucleon re\-scattering we use the current
solution of the GWU/CNS partial wave
analysis~\cite{Strakovsky1,Strakovsky2}. We account for isospin and topological
factors following the prescription given in\\ Refs.~\cite{Locher2,Fasano1,Locher3}.
The phase-space integration is done by the Monte-Carlo method based on
event-by-event simulations. This allows us to apply kinematical cuts on the
final pion momenta similar to that discussed in Refs. \cite{Riedlberger,Ahmad}
in order to investigate discrepancies between the data at spectator proton
momenta below 250 MeV/c. We will come back to this issue later. 

\begin{figure}[b]
\vspace*{-5mm}
\centerline{\hspace*{4mm}\psfig{file=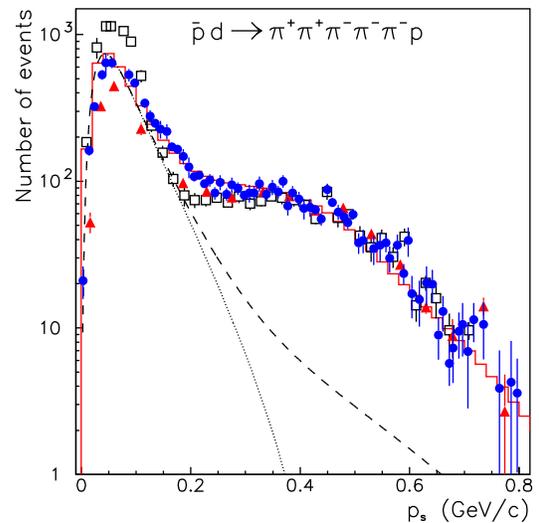,width=8.cm}}
\vspace*{-4mm}
\caption{Proton momentum spectrum for the ${\bar p}d{\to}2\pi^+3\pi^-{p}$
reaction. The data are from Refs.~\cite{Riedlberger} (squares), \cite{Ahmad}
(circles) and \cite{Zemany} (triangles). The dashed line shows the result from
Eq.~(\ref{tree}) taking into account $s$- and $d$-wave parts of the deuteron wave
function  while the dotted line is based on the $s$-wave component alone. 
The solid histogram is the full calculation 
including the nucleon exchange Born diagram and rescattering
diagrams using the GWU/CNS $\pi N$ partial-wave amplitudes~\cite{Strakovsky1,Strakovsky2} 
for $T_M$.} 
\label{bras1}
\end{figure}

Experimental results for the proton momentum distribution for the reaction 
${\bar p}d{\to}2\pi^+3\pi^-{p}$ are shown in Fig.~\ref{bras1}.
The squares \cite{Riedlberger} and circles \cite{Ahmad} are data from the
LEAR facility while the triangles are from an experiment~\cite{Zemany} at 
the Brookhaven National Laboratory (BNL) using the deuterium bubble chamber. 

The basic difficulty in the counter measurements~\cite{Riedlberger,Ahmad}
is the reconstruction of the low-momentum part of the spectator spectra. 
As mentioned in Ref.~\cite{Riedlberger}, for the direct proton detection at 
least transverse proton momenta of $130$~MeV/c are required. 
The very low momentum protons can be only reconstructed through exclusive
measurements of the final pions and applying the missing momentum method. 
But such a reconstruction introduces additional uncertainties in the 
low momentum spectator spectra. To avoid any ambiguity we normalized
all data sets at momenta around 400~MeV/c, i.e. at values where the
proton spectator momentum was measured directly. 
This normalization emphasizes that the shape of the measured spectrum  
from the different experiments is almost identical at higher spectator 
proton momenta. On the other hand, 
we see a substantial disagreement between the available
data~\cite{Riedlberger,Ahmad,Zemany} at momenta below $\simeq 250$~MeV/c.

\begin{figure}[t]
\vspace*{-6mm}
\centerline{\hspace*{3mm}
\psfig{file=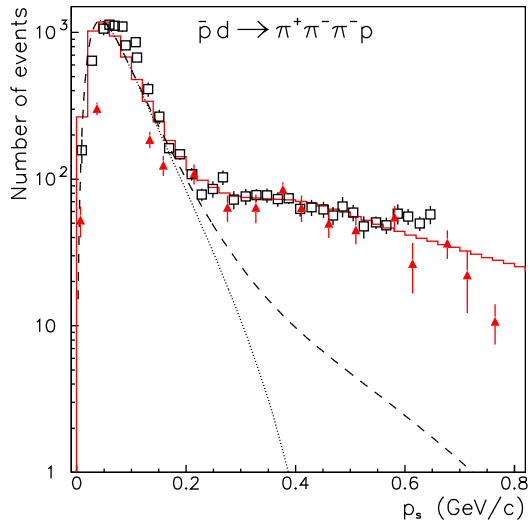,width=8.cm}}
\vspace*{-4mm}
\caption{Proton momentum spectrum for the ${\bar p}d{\to}\pi^+2\pi^-{p}$
reaction. The data are from Ref.~\cite{Riedlberger} (squares) and \cite{Zemany}
(triangles). Same description of curves as in Fig.~\ref{bras1}. 
}
\label{bras2}
\end{figure}

Experimental results for the reaction ${\bar p}d{\to}\pi^+2\pi^-{p}$ are presented 
in Fig.~\ref{bras2}. Again, the squares are LEAR data~\cite{Riedlberger}
while the triangles are from the BNL~\cite{Zemany}. Both data sets are
normalized at proton momenta around 400 MeV/c and in such a way that the 
scale is roughly the same as in Fig.~\ref{bras1}. This allows us to compare
the shapes of the spectator proton distributions for the two reactions. 
There is clearly a difference between those shapes for 
${\bar p}d{\to}2\pi^+3\pi^-{p}$ and ${\bar p}d{\to}\pi^+2\pi^-{p}$. 
Indeed one expects that the shape of the spectator momentum distribution 
depends on the momentum carried by the scattering meson and, consequently, 
that reactions with different final pion multiplicity would exhibit 
different shapes of the proton spectra. The
more energetic pions from the ${\bar p}d{\to}\pi^+2\pi^-{p}$ reaction 
produce more energetic spectator protons in the rescattering.
Qualitatively this follows from the rescattering amplitude of
Eq.~(\ref{rescat}).

Again, for ${\bar p}d{\to}\pi^+2\pi^-{p}$ the experimental results are 
consistent at proton momenta above $250$~MeV/c and disagree substantially at 
low momenta. As already indicated above, we applied kinematical cuts 
on the final pion momenta, similar to those discussed 
in Refs.~\cite{Riedlberger,Ahmad}, in the course of our investigation 
for the reaction ${\bar p}d{\to}\pi^+2\pi^-{p}$ as well as for 
${\bar p}d{\to}2\pi^+3\pi^-{p}$.
But it turned out that those cuts do not resolve the
disagreement between the data~\cite{Riedlberger,Ahmad,Zemany} at spectator
proton momenta below $250$~MeV/c.

Let us now come to the results of our model calculation and first discuss
the normalization, which is a somewhat delicate issue. From a theoretical
point of view it should be done preferrably around the peak in the distribution 
at low momenta where the spectrum is dominated by the Born diagram and 
the $s$-wave component of the deuteron wave function. 
But this is exactly the region 
where the experimental uncertainty is very large. Thus, we decided to 
normalize our results also in the plateau region, i.e. around 400 MeV/c.
Note that the normalization is done for the full calculation. The relative
size of the momentum distribution at the peak as compared to the plateau is fixed
by the ingredients of the model alone, i.e. the contributions of the Born diagram
and from the rescattering diagram. There is no additional normalization 
constant involved here. 
 
The dashed lines in Figs.~\ref{bras1} and \ref{bras2} correspond to the 
contribution from the nucleon-exchange Born diagram given by Eq.~(\ref{tree}). 
Interestingly, our predictions for low momenta agree well with the data 
of \cite{Ahmad} (for ${\bar p}d{\to}2\pi^+3\pi^-{p}$) and 
\cite{Riedlberger} (for ${\bar p}d{\to}\pi^+2\pi^-{p}$). 
For proton momenta above 200 MeV/c all data show a substantial enhancement 
with respect to the predictions based on the Born term alone. For
illustration purposes we show here also results using only 
the $s$-wave part of the deuteron wave function (dotted line).
There have been some speculations that the enhancement at higher momenta 
could indicate an excitation of the short range component of the deuteron 
wave function~\cite{Benz,Mulders,Kim1,Glozman,Lu}. But, in any case, 
a simple renormalization of the $d$-wave contribution would not reproduce the
observed shape of the spectator proton distribution around the plateau, i.e.
for $p_s{\simeq}$400~MeV/c.

The solid histograms in Figs.~ \ref{bras1} and \ref{bras2} are the results of
our full calculation including the nucleon-exchange Born diagram and the
rescattering diagram. But we should say that the off-shell corrections 
of Eq.~(\ref{off}), not considered in the present investigation, are known 
to lead to variations of the order of $30$\% or more in the absolute value of 
the rescattering contribution, though they do not effect the shape of the 
spectator proton distribution \cite{Fasano1}.

Our results are in reasonable agreement with that of Ref.~\cite{Fasano1}, but
are in contradiction to the conclusions of Ref.~\cite{Kudryavtsev}, where the
spectator proton momentum distribution is well reproduced by taking into account
the Born diagram of Fig.~\ref{diag}a) and pion absorption on the spectator 
nucleon. However, there is strong experimental evidence in favor of the
rescattering mechanism: The invariant mass
spectrum of the $2\pi^+2\pi^-$ system from the ${\bar p}d{\to}2\pi^+3\pi^-{p}$
reaction was measured~\cite{Ahmad} for different cuts on the spectator proton
momenta. If the final pions do not undergo rescattering, such cuts should not 
change the invariant mass distribution. But in the experiment
it turned out that the $2\pi^+2\pi^-$ invariant mass distribution depends
substantially on the proton momentum cut when taken below or above
$p_s~{=}~200$~MeV/c.


\begin{thebibliography}{20}
\bibitem{Matsui:1986dk}
         T.~Matsui and H.~Satz, 
         %``J/psi Suppression by Quark-Gluon Plasma
	 %Formation,'' 
         Phys.\ Lett.\  B {\bf 178}, 416 (1986). 
\bibitem{Satz:2005hx}
	 H.~Satz,
	 %``Colour deconfinement and quarkonium binding,''
	 J.\ Phys.\ G {\bf 32}, R25 (2006)
	 [arXiv:hep-ph/0512217].
\bibitem{Batsouli}
	 S. Batsouli, S. Kelly, M. Gyulassy and J.L. Nagle, Phys. Lett. B
	 {\bf 557}, 26 (2003) [arXiv:nucl-th/0212068].
\bibitem{Djordjevic1}
	 M. Djordjevic and M. Gyulassy, Phys. Lett. B {\bf 560}, 37 (2003)
[arXiv:nucl-th/0302069].
\bibitem{Armesto}
	N. Armesto, C.A. Salgado and U.A. Wiedemann, Phys. Rev. D {\bf 69},
114003 (2004)  [arXiv:hep-ph/0312106].
\bibitem{Djordjevic2}
	M. Djordjevic and M. Gyulassy, Nucl. Phys. A {\bf 733}, 265 (2004)
[arXiv:nucl-th/0310076].
\bibitem{PDG}
	W.-M. Yao {\it et al.} [Particle Data Group],   J. Phys. G
	{\bf 33}, 1 (2006)
\bibitem{Moore}
	G.D. Moore and D. Teaney, Phys. Rev. C {\bf 71}, 064904 (2005)
	[arXiv:hep-ph/0412346].
\bibitem{Hees1}
	H. van Hees and R. Rapp, Phys. Rev. C {\bf 71}, 034907 (2005)
	[arXiv:nucl-th/0412015].
	\bibitem{Hees2}
	H. van Hees, V. Greco and R. Rapp, Phys. Rev. C {\bf 73},
	034913 (2006) [arXiv:nucl-th/0508055
\bibitem{Wicks}
	S. Wicks, M. Djordjevic, C. Horowitz and M. Gyulassy, Nucl.
	Phys. A {\bf 784}, 426 (2007) [arXiv:nucl-th/0512076].
\bibitem{Adler}
	S.S.Adler {\it et al.} [PHENIX Collaboration], Phys. Rev. Lett.
	{\bf 94}, 082301 (2005)[arXiv:nucl-ex/0409028].
\bibitem{Bielcik}
	J. Bielcik {\it et al.} [STAR Collaboration],  Nucl. Phys. A
	{\bf 774}, 697 (2006) [arXiv:nucl-ex/0511005].
\bibitem{Zhang}
	Y. Zhang {\it et al.} [STAR Collaboration],  J. Phys. G {\bf
	32}, S529 (2006) [arXiv:nucl-ex/0607011].
\bibitem{Zhong}
	 C. Zhong {\it et al.} [STAR Collaboration],  J. Phys. G {\bf
	34}, S741 (2007) [arXiv:nucl-ex/0702014].
\bibitem{Panda} M. Kotulla et al.,
	{\it Technical Progress Report for $\bar{P}$ANDA, Strong
	Interaction  Studies with Antiprotons}, February 2005, {\rm
	http://www-panda.gsi.de/db/papersDB/PC19-050217\_panda\_tpr.pdf}.
\bibitem{Hartmann}
	O.N. Hartmann {\it et al.} [PANDA Collaboration], Int. J.
	Mod. Phys. A {\bf 22}, 578 (2007).
\bibitem{Brinkmann}
	K.T. Brinkmann, P. Gianotti, I. Lehmann [PANDA
	Collaboration], Nucl. Phys. News {\bf 16}, 15 (2006)
	[arXiv:physics/0701090].
\bibitem{Kuhn}
	W. Kuhn [PANDA Collaboration], Acta Phys. Polon. B {\bf 37},
	129 (2006).
%%%%%%%%%%%%%%%%%%%%%%%%

\bibitem{Haidenbauer}
	J. Haidenbauer, G. Krein, Ulf-G. Mei{\ss}ner and  A. Sibirtsev,
	Eur. Phys. J. A {\bf 33}, 107 (2007)  [arXiv:nucl-th/0704.3668].
\bibitem{DN} 
	J.~Haidenbauer et al., in preparation. 
	\bibitem{Sibirtsev:1999js}
	A.~Sibirtsev, K.~Tsushima and A.~W.~Thomas,
	%``On studying charm in nuclei through antiproton annihilation,''
	Eur.\ Phys.\ J.\ A {\bf 6}, 351 (1999)
	[arXiv:nucl-th/9904016].
\bibitem{Cassing:1999wp}
	W.~Cassing, Y.~S.~Golubeva and L.~A.~Kondratyuk,
	%``Interactions of charmed mesons with nucleons in the anti-p d
	%reaction,''
	Eur.\ Phys.\ J.\ A {\bf 7}, 279 (2000)
	[arXiv:nucl-th/9911026].
\bibitem{Sib01}
	A.~Sibirtsev,
	%``Interactions of charmed mesons with nucleons,''
	Nucl.\ Phys.\ A {\bf 680}, 274c (2001).
%%%%%%%%%% new references
	\bibitem{Locher0}
	C.G. Fasano and M.P. Locher,  Z. Phys. A {\bf 336}, 469
	(1990).
\bibitem{Kolybasov1}
	V.M. Kolybasov, I.S. Shapiro and Yu.N. Sokolskikh, Phys.
	Lett. B {\bf 222}, 135 (1989).
\bibitem{Laget1}
	J.M. Laget, Phys. Rept. {\bf 69}, 1 (1981).
\bibitem{Fasano2}
	C.G. Fasano and M.P. Locher,  Z. Phys.  A {\bf
	338}, 197 (1991).
\bibitem{Laget2}
	J.M. Laget, Nucl. Phys. A {\bf 296}, 388 (1978).
\bibitem{Locher2}
	M.P. Locher and B.S. Zou, Z. Phys. A {\bf 340},  187 (1991).
\bibitem{Fasano1}
	C.G. Fasano, M.P. Locher and S. Nozawa,  Z. Phys.  A {\bf
	338}, 95 (1991).
\bibitem{Voronov}
	D.V. Voronov and V.M. Kolybasov,  JETP Lett. {\bf 57}, 162
	(1993).
\bibitem{Byckling}
	E. Byckling and K. Kajantie, Particle Kinematics (Wiley and
	Sons, New York, 1973).
\bibitem{Laget3}
	J.M. Laget, Phys. Rev. C {\bf 73}, 044003 (2006).
\bibitem{Locher3}
	M.P. Locher and B.S. Zou, J. Phys. G {\bf 19}, 463 (1993).
\bibitem{Laget4}
	J.M. Laget,  Phys. Rev. C {\bf 75}, 014002 (2007)
	[arXiv:nucl-th/0603009].
\bibitem{Machleidt}
	R. Machleidt,  Phys. Rev. C {\bf 63}, 024001 (2001)
	[arXiv:nucl-th/0006014].
\bibitem{Lacombe}
  M.~Lacombe, B.~Loiseau, J.~M.~Richard, R.~Vinh Mau, J.~Cote, P.~Pires and R.~De Tourreil,
  %``Parametrization Of The Paris N N Potential,''
  Phys.\ Rev.\  C {\bf 21}, 861 (1980).
\bibitem{MHE}
  R. Machleidt, K. Holinde, and Ch. Elster, Phys. Rept. {\bf 149}, 1 (1987).
\bibitem{Lin00}
	Z.~w.~Lin, C.~M.~Ko and B.~Zhang,
	%``Hadronic scattering of charm mesons,''
	Phys.\ Rev.\ C {\bf 61}, 024904 (2000)
	[arXiv:nucl-th/9905003].
\bibitem{Hofmann05}
	J.~Hofmann and M.~F.~M.~Lutz,
	%``Coupled-channel study of crypto-exotic baryons with charm,''
	Nucl.\ Phys.\ A {\bf 763}, 90 (2005)
	[arXiv:hep-ph/0507071].
\bibitem{Tolos04}
	L.~Tolos, J.~Schaffner-Bielich and A.~Mishra,
	%``Properties of D-mesons in nuclear matter within a self-consistent
	%coupled-channel approach,''
	Phys.\ Rev.\ C {\bf 70}, 025203 (2004)
	[arXiv:nucl-th/0404064].
\bibitem{Lutz06}
	M.~F.~M.~Lutz and C.~L.~Korpa,
	%``Open-charm systems in cold nuclear matter,''
	Phys.\ Lett.\  B {\bf 633}, 43 (2006)
	[arXiv:nucl-th/0510006].
\bibitem{Mizutani06}
	T.~Mizutani and A.~Ramos,
	%``D mesons in nuclear matter: A D N coupled-channel equations
	%approach,'' 
	Phys.\ Rev.\  C {\bf 74}, 065201 (2006)
	[arXiv:hep-ph/0607257].
\bibitem{Juel1} 
	R. B{\"u}ttgen, K. Holinde, A. M{\"u}ller--Groeling,
	J. Speth, and P. Wyborny, Nucl. Phys. A {\bf 506}, 586 (1990).
\bibitem{Juel2} 
	M. Hoffmann, J.W. Durso, K. Holinde, B.C. Pearce,
	and J. Speth, Nucl. Phys. A {\bf 593}, 341 (1995).
\bibitem{HHK}
	D.~Hadjimichef, J.~Haidenbauer and G.~Krein,
	%``Short-range repulsion and isospin dependence in the K N system,''
	Phys.\ Rev.\ C {\bf 66}, 055214 (2002)
	[arXiv:nucl-th/0209026].
	%%CITATION = PHRVA,C66,055214;%%
\bibitem{Juel3} 
	J. Haidenbauer and G. Krein,
	%``Influence of a Z(1540)+ resonance on K+ N scattering,''
	Phys.\ Rev.\ C {\bf 68}, 052201 (2003)
	[arXiv:hep-ph/0309243].
\bibitem{Tolos08}
	L.~Tolos, A.~Ramos and T.~Mizutani,
	%``Open charm in nuclear matter at finite temperature,''
	Phys.\ Rev.\ C {\bf 77}, 015207 (2008)
	[arXiv:0710.2684 [nucl-th]].
	%%%%%%%%%%%%%%%%%%%%%%%%%%%%%%%%%%%%%%%%%%%%
\bibitem{MG}
	A.~M\"uller-Groehling, K. Holinde, and J. Speth,
	Nucl. Phys. A {\bf 513}, 557 (1990).
	%%%%%%%%%% new references
\bibitem{Lutz04}
	M.~F.~M.~Lutz and E.~E.~Kolomeitsev,
	%``On charm baryon resonances and chiral symmetry,''
 	Nucl.\ Phys.\  A {\bf 730}, 110 (2004)
  	[arXiv:hep-ph/0307233].
\bibitem{Krehl}
	O.~Krehl, C.~Hanhart, S.~Krewald and J.~Speth,
  	%``What is the structure of the Roper resonance?,''
  	Phys.\ Rev.\ C {\bf 62}, 025207 (2000)
  	[arXiv:nucl-th/9911080].
\bibitem{Achot}
  	A.~M.~Gasparyan, J.~Haidenbauer, C.~Hanhart and J.~Speth,
  	%``Pion nucleon scattering in a meson exchange model,''
  	Phys.\ Rev.\  C {\bf 68}, 045207 (2003)
  	[arXiv:nucl-th/0307072].
	%%%%%%%%%%%%%%%%%%%%%%%%%%%%%%%%%%%%%%%%%%%%
\bibitem{Kroll89}
  P.~Kroll, B.~Quadder and W.~Schweiger,
  %``EXCLUSIVE PRODUCTION OF HEAVY FLAVORS IN PROTON - ANTI-PROTON
  %ANNIHILATION,''
  Nucl.\ Phys.\  B {\bf 316}, 373 (1989).
\bibitem{Kai94}
  A.~B.~Kaidalov and P.~E.~Volkovitsky,
  %``Binary Reactions In Anti-P P Collisions At
  %Intermediate-Energies,''
  Z.\ Phys.\  C {\bf 63}, 517 (1994).
\bibitem{Kerbi95}
  B.~Kerbikov and D.~Kharzeev,
  %``N anti-N annihilation at the open charm threshold,''
  Phys.\ Rev.\  D {\bf 51}, 6103 (1995)
  [arXiv:hep-ph/9408378].
\bibitem{Schweiger} W. Schweiger, private communication. 
\bibitem{Hanhart}
	C. Hanhart, Phys. Rept. {\bf 397}, 155 (2004)
	[arXiv:hep-ph/0311341].
\bibitem{SibirtsevK}
	A. Sibirtsev, J. Haidenbauer, H.-W. Hammer and
	S. Krewald, Eur. Phys. J. A {\bf 27}, 269 (2006)
	[arXiv:nucl-th/0512059].
\bibitem{SibirtsevK1}
	A. Sibirtsev, J. Haidenbauer, U.-G. Mei{\ss}ner
	Phys. Rev. Lett. {\bf 98}, 039101 (2007)
	[arXiv:hep-ph/0607212].
\bibitem{TOF}
	S. Abdel-Samad {\it et al.} [COSY-TOF Collaboration] Phys. Lett. B {\bf
	632}, 27 (2006).
\bibitem{SibirtsevAF}
	A. Sibirtsev, K. Tsushima and A. Faessler, Z. Phys. A {\bf 354}, 215
	(1996) [arXiv:nucl-th/9511008].
\bibitem{Barmin}
	V.V. Barmin {\it et al.} [DIANA Collaboration], Nucl. Phys. A
	{\bf 683}, 305 (2001).
\bibitem{ANKE}
	M. Hartmann, Yu. Kiselev {\it et al.} [ANKE Collaboration],  COSY
	Proposal {\bf 147}.
\bibitem{Golubeva}
	Y.S. Golubeva, W. Cassing, L.A. Kondratyuk, A. Sibirtsev and
	M. B{\"u}scher, Eur. Phys. J. A {\bf 7}, (2000)
	[arXiv:nucl-th/9910028].
\bibitem{Riedlberger}
	 J. Riedlberger {\it et al.}, Phys. Rev. C {\bf 40}, 2717
	(1989).
\bibitem{Ahmad}
	S. Ahmad {\it et al.}, IV$^{th}$ LEAR Workshop: Physics at LEAR with
        Low-Energy Antiprotons, Villars-sur-Ollon, Switzerland (1987), 
        edited by C. Amsler {\it et al.},
        Nucl. Sci. Res. Conf. Ser. {\bf 14}, 447 (1988). 
\bibitem{Zemany}
	P.D. Zemany, Z.M. Ma and J.M. Mountz, Phys. Rev. Lett. {\bf 38},
	1443 (1977).
\bibitem{Vandermeulen1}
	J. Vandermeulen, Z. Phys. C {\bf 37}, 563 (1988).
\bibitem{Amsler}
	C. Amsler, Rev. Mod. Phys. {\bf 70}, 1293 (1998)
	[arXiv:hep-ex/9708025].
\bibitem{Strakovsky1}
	R.A. Arndt, I.I. Strakovsky and  R.L. Workman
	Phys. Rev. C {\bf 52}, 2120 (1995) [arXiv:nucl-th/9505040].
\bibitem{Strakovsky2}
	R.A. Arndt, W.J. Briscoe, I.I. Strakovsky and R.L. Workman,
	Phys. Rev. C {\bf 74}, 045205 (2006)
	[arXiv:nucl-th/0605082].
\bibitem{Benz}
	P. Benz and P. S{\"o}ding, Phys. Lett. B {\bf 52}, 367 (1974).
\bibitem{Mulders}
	P.J. Mulders  and  A.W. Thomas, Phys. Rev. Lett. {\bf 52},
	1199 (1984).
\bibitem{Kim1}
	Y.E. Kim and M. Orlowski, Phys. Rev. C {\bf 29}, 2299 (1984).
\bibitem{Glozman}
	L.Ya. Glozman, Prog. Part. Nucl. Phys. {\bf 34} 123 (1995).
\bibitem{Lu}
	L.-C.Lu and  T.-S. Cheng, Phys. Lett. B {\bf 386}, 69 (1996).
\bibitem{Kudryavtsev}
	A.E. Kudryavtsev and  V.E. Tarasov, Sov. J. Nucl. Phys.
	{\bf 54}, 36 (1991).
%%%%%%%%%%%%%%%%%%%%%%%%%%%%%
\end{thebibliography}
\end{document}